\documentstyle[twocolumn,graphicx,epsfig]{mn}

\newif\ifAMStwofonts
\ifoldfss
  \ifCUPmtlplainloaded \else
    \NewTextAlphabet{textbfit} {cmbxti10} {}
    \NewTextAlphabet{textbfss} {cmssbx10} {}
    \NewMathAlphabet{mathbfit} {cmbxti10} {} % for math mode
    \NewMathAlphabet{mathbfss} {cmssbx10} {} %  "   "    "
  \fi
  \ifAMStwofonts
    \ifCUPmtlplainloaded \else
      \NewSymbolFont{upmath} {eurm10}
      \NewSymbolFont{AMSa} {msam10}
      \NewMathSymbol{\upi}     {0}{upmath}{19}
      \NewMathSymbol{\umu}     {0}{upmath}{16}
      \NewMathSymbol{\upartial}{0}{upmath}{40}
      \NewMathSymbol{\leqslant}{3}{AMSa}{36}
      \NewMathSymbol{\geqslant}{3}{AMSa}{3E}

    \fi
  \fi
\fi % End of OFSS

\ifnfssone
  \newmathalphabet{\mathit}
  \addtoversion{normal}{\mathit}{cmr}{m}{it}
  \addtoversion{bold}{\mathit}{cmr}{bx}{it}
  \newmathalphabet{\mathbfit} % math mode version of \textbfit{..}
  \addtoversion{normal}{\mathbfit}{cmr}{bx}{it}
  \addtoversion{bold}{\mathbfit}{cmr}{bx}{it}
  \newmathalphabet{\mathbfss} % math mode version of \textbfss{..}
  \addtoversion{normal}{\mathbfss}{cmss}{bx}{n}
  \addtoversion{bold}{\mathbfss}{cmss}{bx}{n}
  \ifAMStwofonts
    \ifCUPmtlplainloaded \else
      \UseAMStwoboldmath
      \makeatletter
      \new@mathgroup\upmath@group
      \define@mathgroup\mv@normal\upmath@group{eur}{m}{n}
      \define@mathgroup\mv@bold\upmath@group{eur}{b}{n}
      \edef\UPM{\hexnumber\upmath@group}
      \new@mathgroup\amsa@group
      \define@mathgroup\mv@normal\amsa@group{msa}{m}{n}
      \define@mathgroup\mv@bold\amsa@group{msa}{m}{n}
      \edef\AMSa{\hexnumber\amsa@group}
      \makeatother
      \mathchardef\upi="0\UPM19
      \mathchardef\umu="0\UPM16
      \mathchardef\upartial="0\UPM40
      \mathchardef\leqslant="3\AMSa36
      \mathchardef\geqslant="3\AMSa3E
    \fi
  \fi
\fi % End of NFSS release 1

\ifnfsstwo
  \DeclareMathAlphabet{\mathbfit}{OT1}{cmr}{bx}{it}
  \SetMathAlphabet\mathbfit{bold}{OT1}{cmr}{bx}{it}
  \DeclareMathAlphabet{\mathbfss}{OT1}{cmss}{bx}{n}
  \SetMathAlphabet\mathbfss{bold}{OT1}{cmss}{bx}{n}
  \ifAMStwofonts
    \ifCUPmtlplainloaded \else
      \DeclareSymbolFont{UPM}{U}{eur}{m}{n}
      \SetSymbolFont{UPM}{bold}{U}{eur}{b}{n}
      \DeclareSymbolFont{AMSa}{U}{msa}{m}{n}
      \DeclareMathSymbol{\upi}{0}{UPM}{"19}
      \DeclareMathSymbol{\umu}{0}{UPM}{"16}
      \DeclareMathSymbol{\upartial}{0}{UPM}{"40}
      \DeclareMathSymbol{\leqslant}{3}{AMSa}{"36}
      \DeclareMathSymbol{\geqslant}{3}{AMSa}{"3E}
    \fi
  \fi
\fi % End of NFSS release 2

\ifCUPmtlplainloaded \else
  \ifAMStwofonts \else % If no AMS fonts
    \def\upi{\pi}
    \def\umu{\mu}
    \def\upartial{\partial}
  \fi
\fi

\title{The Halo Density Profiles  with  Non-Standard \\
       N-body simulations}
\author[E. D'Onghia$^{1,2}$, C. Firmani$^{3,4}$, G. Chincarini$^{3,5}$]
        {E.   D'Onghia$^{1,2}$, C. Firmani$^{3,4}$, G. Chincarini$^{3,5}$\\
        1- Universita' degli Studi di Milano, via Celoria 16, Milano, Italy\\
        2- Max-Planck-Institut f\"ur Astronomie, K\"onigstuhl 17, 69117 Heidelberg, Germany\\
        3- Osservatorio Astronomico di Brera-Merate, via Bianchi 46, 23807 Merate (LC), Italy\\
        4- Instituto de Astronomia, UNAM, A.P. 70-264, 04510 Mexico D.F., Mexico\\
        5- Universita' degli Studi di Milano-Bicocca, Piazza dell'Ateneo Nuovo 1, 20126 Milano, Italy}
       
\date{14 March 2002}

\pagerange{\pageref{firstpage}--\pageref{lastpage}}
\pubyear{2002}

\begin{document}

\maketitle

\label{firstpage}

\begin{abstract} 
We propose a new numerical procedure to simulate a 
single dark halo of any size 
and mass in a hierarchical framework coupling the 
extended Press-Schechter formalism (EPSF) 
to N-body simulations. The procedure consists of assigning
cosmological initial conditions to the particles of a single halo 
with a EPSF technique and following only the
dynamical evolution using a serial N-body code. The computational
box is fixed with a side of  $0.5 h^{-1}$ Mpc. This allows to simulate
galaxy cluster halos using appropriate scaling relations, 
 to ensure savings in computing time and code speed.
The code can  describe the properties of halos composed of collisionless or
collisional dark matter.
For collisionless Cold Dark Matter (CDM) particles the NFW profile is reproduced
for galactic halos as well as galaxy cluster halos.
Using this numerical technique we study some characteristics of halos 
assumed to be isolated or placed 
in a cosmological context in presence of weak self-interacting
dark matter: the soft core formation and the core collapse.
The self-interacting dark matter
cross section per unit mass is assumed to be inversely proportional to the particle collision 
velocity: $\sigma/m_{x} \propto 1/v$.

\end{abstract}


\section{Introduction}

A detailed understanding of structure formation 
is one of the central goals of contemporary astrophysics
and cosmology. In the popular hierarchical clustering framework 
(White $\&$ Reese 1978) luminous galaxies form by
gas cooling and condensing within dark matter halos. These
halos merge to build larger structures, hierarchically. 
In the CDM  scenario  
systematic studies of halo density profiles
for a wide range of halo masses were derived by
Navarro, Frenk $\&$ White (1997; hereafter NFW) who argued 
that the analytical profile of the form: $\rho(r)=\rho_s(r/r_s)^{-1}
(1+r/r_s)^{-2}$ provides a good description of halo profiles 
for all halo masses, where $r_s$ is the scale radius which corresponds
to the scale at which the slope of the profile is $-2$. 

While predictions of the CDM models have  successfully 
accounted for many observations at large scale scales, 
increasing interest in testing the predictions
at subgalactic and galactic scales
has grown in the last few years.  
Interest began from indications that observed HI
rotation curves  in the central regions of dark matter dominated dwarf 
and Low Surface Brightness (LSB) galaxies are at odds with predictions 
from hierarchical models
(Flores $\&$ Primack 1994; Moore 1994; Burkert 1995). The  
 models predict circular velocities that increase too rapidly 
with growing radius compared to the observed profiles of rotation curves. This 
fact 
implies a failure of the shape of the predicted halo density
profile.
Recent high-resolution N-body simulations have shown that, as the numerical 
resolution is increased, the inner profiles is $\rho \propto r^{-1.5}$, 
even steeper than the NFW profile  (Moore et al. 1999; Fukushige $\&$ Makino 2001),
exacerbating even more the discrepancy with the observations. 
On the other hand, Swaters, Madore $\&$ Trewella (2000), providing 
H$\alpha$ rotation curves for some LSB galaxies, challenged the existence of
soft cores for these galaxies. They pointed out that the HI 
rotation curves are affected by poor spatial resolution, causing a 
smoothing of the curves in the inner regions (usually named beam smearing).
For a sample of late-type dwarf galaxies, 
van den Bosch $\&$ Swaters (2001) 
show that  beam smearing
does not allow for distinguishing between the existence  or lack of 
soft cores in late-type dwarf galaxies. However, high-resolution
H$\alpha$ rotation curves of LSB galaxies of de Blok, McGaugh $\&$ Rubin (2001)
and Marchesini et al. (2002) are in favour of core-dominated halos.
Similar results are derived by Salucci $\&$ Burkert (2000).
While the question of the existence  or lack
of shallow cores in galaxies continues to be debated,  some authors have 
revealed the presence of soft cores
at the centre of some clusters of galaxies  
from strong lensing observations in Cl0024+1654 (Tyson, Kochansky $\&$
Dell'Antonio 1998) and from  X-ray data in Abell 1795 (Ettori $\&$ Fabian 2002). 
Intriguely, clusters showing a soft core do not have a  
prominent central cD galaxy. In particular 
Abell 1795 has a central cD galaxy moving far from the centre (Ettori $\&$ Fabian 2002). 
%However,  the presence of shallow cores remains highly
%uncertain in clusters of galaxies, because of the small number of
%cluster observed. 
On the other hand, Chandra X-ray data show the 
lack of any core at the centre of HydraA (David et al. 2001) and a 
very small core in EMSS 1358+6245 (Arabadjis et al. 2002); 
both clusters have a central cD galaxy.
In relaxed clusters, a dominant central galaxy could play a role 
making steeper the primordial halo density profile. 
Accepting the presence of soft cores advocated in Abell 1795 and Cl0024+1654,  
Firmani and co-workers (2001) have correlated the halo scales from galaxies
to galaxy clusters, integrating the available information. Surprisingly, they
find that, in this case,  the central density of dark halos 
is independent of the halo mass
with a value of $0.05 h^2 \ M_{\odot}$ pc$^{-3}$, and that the core radius 
increases proportionally with the halo maximum rotation velocity. 

The presence or the lack of soft cores at the centre of galaxy clusters
has interesting consequences for the nature of the dark matter.
The lack of any core in clusters makes attractive  several
scenarios in the debate over the nature of the dark matter. In warm dark matter models,
the maximum phase space density of the particles
defined as $f_{\rm{max}}\equiv \rho_0/v^3$, with $\rho_0$ the halo central density
and $v$ the halo velocity dispersion, has a finite value and is preserved due to the Liouville theorem, 
implying an increase of the 
halo central density when the halo mass increases: $\rho_0 \propto v^3$ (Sellwood 2000; Hogan $\&$ Dalcanton 2000).
In this scenario very massive halos of galaxy clusters are predicted to have
high central densities.

A different nature for the dark matter was proposed by Spergel
$\&$ Steinhardt (2000) in order to overcome the soft core question: 
if the dark matter is self-interacting, heat transfer
towards the central regions triggers a thermalization process in the 
dark halos avoiding the formation of a cuspy profile.  The effects 
of weak self-interacting dark matter were investigated using N-body
simulations on isolated halos  
(Burkert 2000; Kochaneck $\&$ White 2000) and on   CDM halos  
(Yoshida et al. 2000; Dav\'{e}  et al. 2001). These simulations
were performed assuming a cross section independent on the relative particle 
velocity. 
Ostriker (2000) and Hennawi and Ostriker (2002) have shown a possible
inconsistency of the collisional scenario: indeed the model would cause an exorbitant
grow on supermassive black holes that imposes a very strict upper limit on
the collision cross section. This conclusion derives from their assumption
that when black holes seeds have been formed (at $z < 20$) the innermost dark halo shown a
NFW density profile. This is not necessarily true, at $z < 20$ collisions have lowered the 
central density to values at which the growth of the black holes seeds becomes negligible. 
The other upper limits pointed out by the same authors and concerning the gravothermal
catastrophe and the galactic halo evaporation in clusters are consistent with the
cross section used in this work.  
On the other hand, several authors have 
pointed out some limitations of the collisional dark matter models. 
Assuming a dark matter cross section independent of the particle 
relative velocity, Miralda-Escud\`{e} (2002) shown that this scenario predicts
cluster cores which are too large and round to be consistent with gravitational lensing data.

This work is focussed on the soft core question and explores the possibility that the 
cross section for  the 
dark matter interaction decreases with velocity $\sigma/m_x \propto 1/v$
as suggested by Firmani et al. 2001 and Wyithe et al. 2001.
The focus here is an exploration of 
halo core properties using N-body techniques in a very weak self-interacting regime.
For haloes within a hierarchical context, the cross section 
value we propose rules out ranges in which the evaporation problem and the core
collapse could be significant. 
We show that if further observational data can confirm evidence of shallow cores
in galaxy cluster halos, then self-interacting dark matter 
with a cross section inversely proportional to the halo dispersion
velocity is capable of predicting the existence of soft cores in dwarf
galaxies as well as in galaxy clusters. 
This paper is organized as follows: in section 2 the numerical
technique is described, in section 3 for collisionless dark matter particles, we compare the cosmological
halo density profiles obtained with our technique to the  
NFW models for different size halos. Section 4 introduces the Monte Carlo method for weakly interacting particle
systems implemented in the code. In sections 5 and 
6, the halo density profiles are investigated in a self-interacting 
dark matter scenario for two different cases:  isolated
halos and cosmological halos.

\section{Methodology}
In all numerical simulations, approximations and compromises
are necessary if the calculation is to be completed within a 
reasonable amount of time. An important choice to make is the effective
resolution of the simulation: increasing the spatial resolution
requires increasing both the number of particles to prevent
two-body effects and the number of timesteps to follow the evolution of smaller
structures. For a given simulated volume of universe, the
number of particles determines the mass resolution. In cosmological
simulations involving only collisionless dark matter the choice of
particle number is often driven by computer memory limitations.
At present large cosmological N-body simulations have reached the 
stage where detailed
structural properties of many dark matter halos can be resolved 
simultaneously making use of codes running in parallel 
 and of a very large and  expensive number of particles. 
We use for our simulations the adaptive, particle-particle,
particle-mesh P$^3$M--SPH code, which is basically  
the serial publicly released version
of HYDRA described in detail by Couchman, Thomas $\&$ Pearce (1995). 
The publicly available serial version of the code does not allow 
for cosmological simulations of large volumes of universe
and the resolution of smaller structures.
With the aim of overcoming 
this difficulty without using supercomputers, we have introduced 
relevant modifications to 
the code in order to satisfy 
the following requirements: {\it i)} the code  has to be reasonably fast, serial, 
 and running on a single processor workstation; {\it ii)} 
for each run the code must follow
the dynamical evolution of a single isolated or cosmological 
halo in a limited volume;
{\it iii)} the code has to be capable of describing the properties of collisional or 
collisionless
dark matter halos of different sizes
and masses from galactic to galaxy cluster scales with an arbitrary
number of particles.

The main change to the code concerns the cosmological initial conditions.
Generally  cosmological simulations with N-body grid codes begin by  
fixing a power spectrum of fluctuations, 
  in accordance with a  cosmological
   model on a uniform grid which
  covers the whole computational box. 
The fluctuation field is perturbed in  Fourier space and the 
 Zel'dovich displacement is assigned to the particles.
The grid defines the lowest level of
resolution of the simulation. Subsequently, simulations are run with
a fixed number of particles and halos are identified in the simulation. 
 Once marked, the 
selected halos are re-simulated with higher mass and
force resolution.
%For each selected halo its virial
%radius is determined and the Lagrangian coordinates are found  
%for each particle. The coordinates are
%used to mark blocks of particles to generate the initial conditions of the
%highest mass resolution (see, e.g., Bertschinger 2001).

Here, we propose simulations with a  different approach.   
Assuming a spherical symmetry, the dark matter initial distribution at z=100
of a single halo is very close to a uniform density distribution in expansion.
A small negative radial density gradient causes each shell to reach the 
maximum expansion at the dynamical time corresponding to its inner average density.
Because of the negative density gradient, the dynamical time is higher in the most
external shells. This fact illustrates why the initial negative density gradient 
is equivalent to a mass aggregation history (MAH) for the halo.
In our case, for a given halo present mass, a MAH is obtained from the EPSF developed 
by Avila-Reese, Firmani and Hern\'{a}ndez (1998; hereafter AFH). Using the dynamical 
time of each shell implicit in the MAH we recover the initial density gradient,
i.e. the initial condition at z=100. The stochastic nature of the density fluctuations
induces a stochastic structure on the MAH. We have limited our analysis on the
average MAH from which a given present halo mass has grown.
This initial condition for the description of the
cosmological evolution of growing halos offers clear
advantages with respect to the simulations of isolated
halos starting with a NFW or a Hearnquist density
profile. However, because the mass accretion is
described in a spherical symmetric frame, in this
scheme the tidal destructive action of strong mergers
on a central dense halo core is not taken into
account. This fact establishes a limit on our approach
that works in the sense to make more difficult the
formation of inner shallow density profiles. Because of our
goal is to explore the capability of self-interaction
to create shallow cores, our results should be
strengthened if a detailed cosmological simulation
is made. Our result will be particularly
correct in the case of dwarf and LSB galaxies which,
being fragile disks and field objects, their MAH could
not be affected by major mergers. In the case of galaxy
clusters our result is purely indicative and obeys to
a completeness criterion. In fact in this case major
mergers may amplify the formation of shallow cores as
well as they may introduce significant deviations from
the spherical symmetry. Taking into account the
previous discussion and considering that our goal is
to analyze the ability of self-interaction to
create shallow cores, we concentrate our attention on
the average MAH. The lack of any stochastical
information from observations makes unnecessary any
analysis which takes into account the stochastical
nature of MAHs.
 
The following steps are taken:
\begin{itemize}
\item  We assign to the halo an initial density profile
   described   by  a collection of concentric shells. For each shell, the MAH
      of the halo gives: the time of maximum expansion 
      $t_{max}$, the radius of the maximum expansion,
      the initial radius at a given redshift, and  
      the cumulative mass within the shell. 
\item   The dynamical evolution
of the halo within a non-linear regime is followed by the N-body code.
\end{itemize}
\noindent
At the beginning of 
our N-body simulations 
the radius of  each particle is interpolated to the initial radius, 
determined by the halo MAH at a cumulative value of the mass. 
We have used different techniques to distribute the particles (the halo mass)
within the shells at the beginning of the simulation, with a grid or with
isotropic random direction. In the first case,  particles are assigned on a spherical
mesh points and 
are originally equally spaced on the mesh, then the 
distance of each particle from the centre is computed  according  
to the halo MAH. In the second case, 
for each particle the radius is 
isotropically assigned using a Monte Carlo routine, while the  magnitude of the radius 
is fixed by the cosmological MAH.
Both methods in assigning the particles at the initial conditions
have limits and advantages. The first method uses a grid
to assign particles and thus it is closer to the
standard assignment used in the original version of the code.
We have used it for the cosmological collisionless and collisional runs. 
However, in some cases, 
a weak filament is seen in the virialized halo. This elongated structure
is an artificial numerical effect, and some tests
have shown that it is present already in the original version of the code.
On the other hand, the second technique prevents the formation
of elongated structures but since the particles are 
assigned with random isotropic directions, they rapidly clump, due to the gravity, 
to form substructures at  very early times
that survive during the virialization of the halo.
These substructures are spurious. Thus, for the cosmological simulations
we prefer to assign  particles to a grid.
The Hubble flow at $z=100$ was assigned to the particles,
as initial velocity. The initial value of z has been chosen with the 
criterion to avoid any spurious formation of soft cores.

%------------------------------------------------------------------
\section{Modelling collisionless CDM halos}
%-----------------------------------------------------------------

The focus of the approach proposed here is on the exploration of  
density profiles of halos of any size and mass.
We show now that by making use of the N-body code with the initial conditions
assigned with our new numerical technique, 
we can reproduce
the NFW density profile for a single collisionless dark matter 
cosmological halo of any mass, after the code is run for a few hours (CPU time) 
on a standard workstation. 

We carried out simulations of galactic halos as well as 
galaxy cluster halos within a flat $\Lambda$CDM universe   
with matter density
$\Omega_m=0.3$, cosmological constant $\Omega_{\Lambda}=0.7$, 
and rms linear 
fluctuation amplitude in 8$h^{-1}$ Mpc spheres of $\sigma_8=1$.
Each simulation followed the trajectories of $N=80,000$ dark matter particles 
within a physical cube of side 0.5 $h^{-1}$  Mpc from z=100 to the present.
The Hubble constant is assumed to be of 65 km s$^{-1}$ Mpc$^{-1}$ ($h$=0.65).
Tests with more particles did not change the results.
One refinement level was introduced in the region of 
highest mass density. 

On galactic scales, for a halo mass of $10^{11}h^{-1}$ M$_{\odot}$ the mass per particle used is 
$m_p= M/80,000=1.25 \cdot 10^6 h^{-1}$ M$_{\odot}$ and 
the softening length is fixed in order to obtain a spatial effective 
resolution of $\approx$ 1 $h^{-1}$ kpc.
On scales of galaxy clusters we have 
carried out simulations of a single halo with  
$M=10^{15} h^{-1}  \ M_{\odot}$ with a mass resolution of $m_p=1.25 \cdot 
10^{10} h^{-1} \ M_{\odot}$
and an effective spatial resolution of   $\approx$ 20 $h^{-1}$ kpc. 

We stress that the cluster simulation
was obtained in the same cube of side 0.5 $h^{-1}$ Mpc. In fact,
 as gravitation is a scale invariant physical process,
a halo of any mass and size
may be simulated in the same fixed volume.
Of course, because of  the hierarchical process of mass accretion, the mass 
aggregation history of a galactic halo is different from the galaxy cluster 
halo merging history. Thus, once the MAH is generated for a halo of  
galaxy cluster size, we can evolve the dark halo by N-body simulations by  
rescaling the 
mass M and the radius R to any mass $M_0$ and initial radius $R_0$ with  the
following relations:

\begin{equation}
M=m M_0,
\end{equation}
\begin{equation}
R=r R_0,
\end{equation}
\begin{equation}
t=t_0 (r^3/m)^{0.5}, 
\end{equation}
\begin{equation}
v^2=v_0^2 \left(m/r \right),
\end{equation}
where $m$, $r$ are scaling parameters. 
This work simulates galaxy cluster halos by rescaling 
the cluster mass to mass $M_0$ of a galactic halo with  $10^{11} \ h^{-1} M_{\odot}$.

Simulated density profiles of collisionless dark matter are shown in the
top panel of Figure 1 for two halos with the following masses: 
$M=10^{11} \ h^{-1} M_{\odot}$ (filled circles) and 
$M=10^{15} \ h^{-1} M_{\odot}$ (open circles). Solid lines are the NFW profiles
for the same halos. The good agreement between the halo density profiles 
obtained with our method and the NFW model is encouraging.
The velocity dispersion radial profiles corresponding to the 
same halos are drawn in the bottom
panel of Figure 1 with filled symbols corresponding to the galactic halo
of $M=10^{11} \ h^{-1} M_{\odot}$ and open symbols to galaxy cluster halo of
$M=10^{15} \ h^{-1} M_{\odot}$, as derived by our simulations. The overlapped solid lines
are the radial velocity dispersion for the NFW model with the same mass.

A more quantitative comparison to the NFW model for the two halos 
is obtained deriving values for
the concentration parameter, defined as the ratio between the virial radius and the 
scale radius $c=r_{vir}/r_s$ (Navarro, Frenk $\&$ White 1997). For the halos of 
$M=10^{11} \ h^{-1} M_{\odot}$ and $M=10^{15} \ h^{-1} M_{\odot}$ we estimate 
c=9.5 and 5, respectively for our realizations. The concentration values of our run are plotted in the Figure 2
as filled points and compared to the mass-concentration  relation of the NFW model (dashed line)
$\footnote{The mass-concentration relation of the NFW model is derived using the program 
kindly made available to the community by J.Navarro.}$. The halos are a little more concentrated
than the average concentration predicted by the NFW model. However, high-resolution N-body
simulations have shown a spread around the average
mass-concentration relation (Wechsler et al. 2002), thus the concetrations we find
in our two realizations are inside the spread.
 
\begin{figure}
\epsfig{file=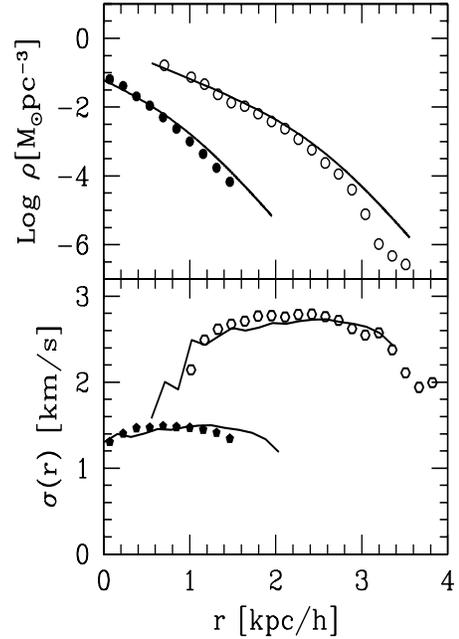,angle=0,width=10cm,height=12cm}
\@ 
\vskip-2cm
\caption{In the top panel simulated density profiles of collisionless 
dark matter are shown using non-standard N-body simulations 
for halos with masses: M$=10^{11} h^{-1} M_{\odot}$ (filled
circles) and  M$=10^{15} h^{-1}  M_{\odot}$ (open circles). A Hubble 
constant of 65 km s$^{-1}$ Mpc$^{-1}$ is adopted. Solid lines
are the NFW profiles for the same halos. Corresponding velocity dispersion 
radial profiles are drawn in the bottom panel with filled symbols corresponding 
to the galactic halo and open symbols to the galaxy cluster halo.
The solid lines are the radial velocity dispersion for the NFW model.}  
\end{figure}

\begin{figure}
\epsfig{file=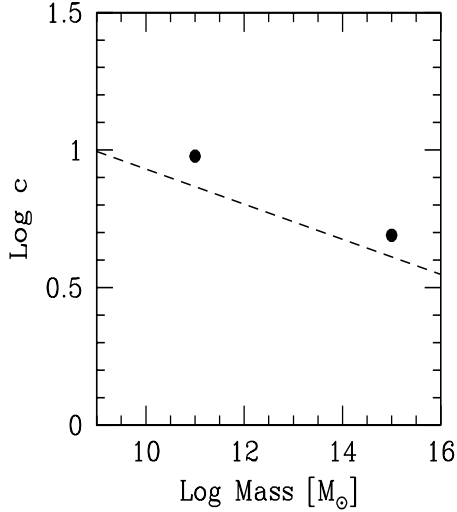,angle=0,width=10cm,height=12cm}
\@
\vskip-2cm
\caption{Concentration parameters $c$ versus the halo mass for the two halos
of M$=10^{11} h^{-1} M_{\odot}$ and  M$=10^{15} h^{-1}  M_{\odot}$ as obtained
from our runs (filled points). The dashed line is the halo Mass-concentration
relation of the NFW model.}
\end{figure}

%-------------------------------------------------
\section{Modelling Collisional CDM Halos} 
%-------------------------------------------------

The idea of a collisional dark matter  suggested by Spergel
$\&$ Steinhardt (2000), is capable of producing soft cores and, at the 
same time, preserves the  success of 
CDM models in explaining the observed
properties of the universe at large scales. 
In fact, in the hierarchical framework of  structure
formation, the halo continuously accretes mass by merging.
However, collisions between particles cause heat transfer inwards. Since
its negative heat capacity, the core gains energy and expands, producing a soft core.
However, the idea of weak self-interacting dark matter poses the following question.
What is the final equilibrium configuration of the halo? The initial
state is a halo described by a NFW model far from thermal equilibrium, 
and the final state is a halo with a thermalized core. 

We explore the dynamical properties of collisional
dark matter studying isolated or cosmological halos of any mass
using N-body techniques. 
We use a Monte Carlo technique to implement the self-interaction 
of the dark matter particles in  collisionless N-body code. 
The scattering process between the dark particles is implemented
with the following algorithm. 
Within a timestep $\Delta t$, the probability $P_i$ of a particle with velocity 
$\vec{v}_i$ to interact with another particle is:

\begin{equation}
P_i=\Gamma_i \ \Delta t,
\label{KWprob}
\end{equation}
with $\Gamma_i$ the scattering rate for a particle with velocity
$v_i$: 
\begin{equation}
\Gamma_i=\Sigma _j^N \Big(\frac{\sigma}{m_x}\Big) 
\frac{m_j [\vec{v}_j-\vec{v}_i]}{V},
\end{equation}
where $\Sigma _j^N m_j/V$ is the estimate of the local density, with
$V=4 \pi r^3/3$.
Although a constant cross section is used by several authors, here we 
assume a cross section inversely proportional
to the particle velocity dispersion as found to be justified
by the observations in the central densities of galaxies and 
galaxy clusters (Firmani et al. 2001; Wyithe et al. 2001)
\begin{equation}
\Big(\frac{\sigma}{m_x}\Big) \cdot v\approx \rm{const} 
\end{equation} 

\noindent
Within a timestep 
the algorithm is as follows: 
\begin{itemize}
\item [{\it i})] For each particle the 
interaction probability $P_i$ is computed.
\item [{\it ii})] For each scattering event
the collision is with one of the nearest particles (N=16, but
increasing N does not change the result). 

\item [{\it iii})] Once a particle is selected for the interaction, the
scattering is isotropic and the diffusion is computed.
\end{itemize}
Concerning the diffusion process, for particles with centre- 
of-mass velocity 
${\vec{v_B}}=({\vec{v_i}}+{\vec{v_j}})/2$ 
and velocity difference  ${\vec{v_0}}=({\vec{v_i}}-{\vec{v_j}})/2$, 
a random direction ${\vec{e}}$
is selected and new velocities are assigned to the colliding 
particles $i$ and $j$
\begin{equation}
{\vec{v_i}}={\vec{v_B}}+\frac{{\vec{e}}}{|e|} |{\vec{v_0}}|,
\end{equation}
\begin{equation}
{\vec{v_j}}={\vec{v_B}}-\frac{{\vec{e}}}{|e|} |{\vec{v_0}}|.
\end{equation}
The energy and linear momentum are preserved.
The numerical method
is similar to the algorithm implemented by Burkert (2000) and Yoshida
et al. (2000) with the
difference that the cross section is dependent upon 
the particle velocity. 
\noindent
For the cross section we assume:
\begin{equation}
\Big(\frac{\sigma}{m_x}\Big)=10^{-24}  
\Big(\frac{100 \ \rm{km} \ \rm{s}^{-1}}{v} \Big) \ \frac{\rm{cm^2}}{\rm{GeV}}.
\end{equation}
In order to simulate a galaxy cluster halo with collisional dark
matter, the cross section value was rescaled as well
as the radius, velocity, time,  and mass according to eqs.(4) to (6) inclusive.

\begin{figure}
\epsfig{file=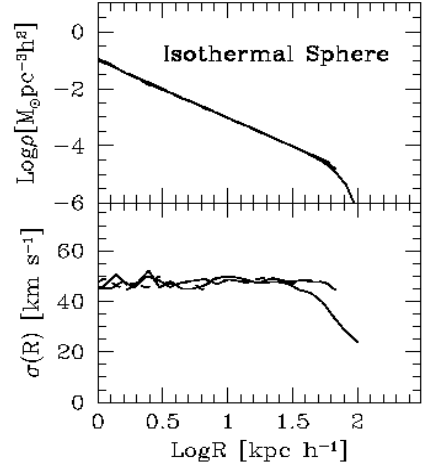,angle=0,width=10cm,height=12cm}
\@ 
\vskip-2cm
\caption{In the top panel the time evolution of the density profile 
is shown for a 
halo of mass M$=10^{11} h^{-1} M_{\odot}$ described by a singular isothermal 
sphere. This run uses collisionless dark matter particles.
After several timescales, the density distribution is represented
by the solid line, while the short dashed line is the profile at the
beginning of the simulation (overlapped). 
In the bottom panel the corresponding radial velocity dispersion is shown.}
\end{figure}

%--------------------------------------------------
\section{Isolated Halos and core collapse}
%-------------------------------------------------------
For isolated halos the balance 
between continuous mass aggregation due to the halo
merging history and the thermalization process induced 
by collisions fails. In this section we investigate whether this case   
can produce core collapse with a consequent gravothermal
catastrophe in a short time.

First, we test if the code is able to preserve the 
state of dynamical equilibrium in a model. We have
simulated a halo with {\it collisionless} dark particles and 
a mass profile of an isothermal 
sphere. This choice is related to the known properties
of this model.
In the top panel of Figure 3, the time
evolution of the density profile is shown for several dynamical
timescales. The density profile preserves the shape for all times.
After several timescales, the density distribution is represented
by the solid line, while the short dashed line is the profile at the
beginning of the simulation. In the bottom panel the corresponding
radial velocity dispersion is shown. 

Note that the isothermal sphere
was settled  by assigning to the particle velocities 
with amplitudes consistent with the Maxwellian velocity distribution
function, $f(v)$ of the
model. In fact, in this case the initial profile is 
at equilibrium and does not show any adjustment 
during dynamical evolution.
The result is remarkable: the dynamical equilibrium of the 
model is preserved for all times.   The only modification to the profile
is after many dynamical timescales in the outer parts 
due to a evaporation process of the particles. 
Furthermore, this model tests the softening radius 
fixed in our simulations.
For a galactic halo of 
$10^{11} \ h^{-1} M_{\odot}$ the softening radius was fixed in order
to obtain a spatial effective resolution of 1 $h^{-1}$ kpc.

\noindent
We have followed the  evolution of dark density profiles 
for a galactic halo of $10^{11} \ h^{-1} M_{\odot}$ formed by  
collisional dark matter and characterized by
the following initial mass distribution: 
the Hernquist model and the King profile.

%--------------------------------------------------------------
\subsection{The Hernquist profile}
%------------------------------------------------------------
An isolated galactic halo model described by an initial Hernquist profile 
(Hernquist 1990) has attracted considerable attention
in the study of self-interacting dark matter.

For this model characterized by a central density cusp ($1/r$) similar
to a NFW profile, Burkert
(2000) and Kochanek $\&$ White (2000)  argue that
soft cores are produced in a short dynamical timescale induced by collisions. 
However the shallow cores so created are unstable
and  a singular isothermal sphere is produced 
as a final halo density profile.

We set up the initial conditions as a Hernquist profile with the
halo mass of $10^{11}  \ h^{-1} M_{\odot}$ and a 
characteristic scale radius $a=50 h^{-1}$  kpc.  
Following the analytical profile, particle positions are  
distributed isotropically and the magnitude of the velocity 
is  computed from the
energy distribution function, $f(E)$, of this model, also assigned
with an isotropic distribution.

\begin{figure}
\epsfig{file=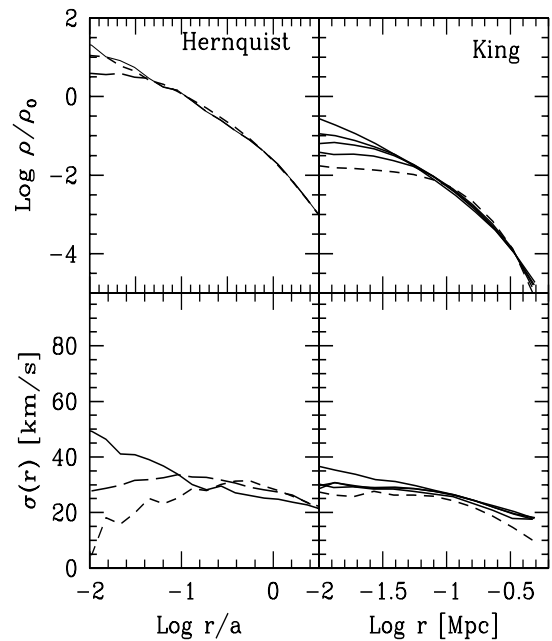,angle=0,width=10cm,height=12cm}
\@
\vskip-2cm
\caption{{\it Left panels.} The time evolution 
for the Hernquist
profile is shown for self-interacting cross section value obtained
assuming a cross section inversely proportional to the halo
dispersion velocity, $(\sigma/m_x)\cdot v_{100}=10^{-24}$ cm$^2$ GeV$^{-1}$,
where $v_{100}$ is the halo dispersion velocity in units of 100 km $s^{-1}$. 
The initial Hernquist profile is shown as a short-dashed line.
After 2.25 dynamical timescales  a minimum central density is reached
(long-dashed lines) and the core collapse after 5 dynamical timescales
 (solid lines). Corresponding velocity dispersion radial profiles
are shown in the bottom. {\it Right panels.} The time evolution of the 
King profile is plotted. Dashed lines are the initial King profile (t=0).
After a while, 
the initial flat core reduces its size towards core collapse.
In the plot the time evolution halo density profile is shown, from 
the lower solid line to the higher solid line representing 
the profile after 0.8 $t_{\rm{dyn}}^K$.
In the bottom panel of the same figure the radial 
velocity dispersion profile is shown for the model.}
\end{figure}

First, we run a model without collisions between particles and we
have shown that the shape of the profile is not modified during
 dynamical evolution. The dynamical equilibrium is preserved.
The dynamical time is defined:
\begin{equation}
t_{\rm{dyn}}= 4\pi \sqrt{\Big(\frac{a^3}{GM}\Big)};
\end{equation}
thus, the dynamical time  is $t_{\rm{dyn}}^H=6.7$ Gyr ($t_{\rm{dyn}}^H$
is referred to the Hernquist model).
\noindent
In the  top-left panel of Figure 4, the time evolution 
for the Hernquist
profile is shown for self-interacting cross section value obtained
assuming a cross section inversely proportional to the halo
dispersion velocity, $(\sigma/m_x)\cdot v_{100}=10^{-24}$ cm$^2$ GeV$^{-1}$,
where $v_{100}$ is the halo dispersion velocity in units of 100 km $s^{-1}$. 
The initial Hernquist profile is shown as a short-dashed line.
After 2.25 dynamical timescales  a minimum central density 
value is reached (long-dashed line), increasing after 4 dynamical timescales   
and reaching a lack of the core after more than 
5 dynamical timescales (solid line),    
due to the collisions between particles.
We also run simulations with very 
high cross section values, finding that
once the soft core is formed, it disappears in a short time,
in agreement with the work of Kochaneck $\&$ White (2000) and Burkert (2000),
when the cross section was adopted to be independent of the particle collision velocity.
Thus, high cross section values seem to work on a catalyst of  the 
core collapse.

%----------------------------------------------------------
\subsection{The King profile}
%------------------------------------------------------------

We have simulated an isolated  halo of $10^{11} \ h^{-1} M_{\odot}$ with a King model 
mass distribution. In fact, the dark density profile of
LSB and dwarf galaxies is well matched by a King profile
with form parameter $P=8$ (Firmani et al. 2001).
We test whether starting with a isothermal configuration and  
a shallow core the isolated halo undergoes core collapse due to
the presence of collisions between the dark particles. We want
to follow the halo core evolution 
through isothermal equilibrium states.
After assigning isotropic
positions to the particles, we pay attention that
at the beginning of the simulation the profile is at
equilibrium, and we calculate the velocity magnitude from the
King  distribution function $f(v)$.
 Again, particle velocities
are isotropically distributed with a Monte Carlo routine. \\

First, we have run the model with $\sigma/m_x=0$. The shape
of the King profile is preserved at all times. 
We now run the model in a strong self-interacting regime:
$(\sigma/m_x)\cdot v_{100} \simeq 5\cdot 10^{-22}$ cm$^2$ GeV$^{-1}$.
 In fact, 
for weak cross sections, we have verified
that this model
evolves slowly towards core collapse. In this case, a large cross
section value is assumed in order to hasten core collapse.  
The dynamical time depends upon the adopted concentration for
 the King model. Defining the dynamical time as above,  
we find that $t_{\rm{dyn}}^K \approx 4 \  t_{\rm{dyn}}^H$
(where $t_{\rm{dyn}}^K$ is referred to the King model and 
$t_{\rm{dyn}}^H$ to the Hernquist model, respectively).
In the top-right panel of Figure 4, the time evolution of the 
King profile is plotted. Dashed lines are the initial King profile (t=0).
After a while, 
the initial flat core reduces its size towards core collapse.
In the plot the time evolution halo density profile is shown, from 
the lower solid line to the higher solid line representing 
the profile after 0.8 $t_{\rm{dyn}}^K$.
In the bottom panel of the same figure the radial 
velocity dispersion profile is shown for the model.

It is interesting to note that any reasonable value for the cross section (weak or
strong) produces core collapse. Weaker  
collisional regimes  will produce core collapse in a longer time.
Of interest is the evidence that a dark halo described by a King 
profile evolves via equilibrium states
towards the gravothermal catastrophe under the effect of collisions.

%---------------------------------------------------------------
\section{Collisional halos in a hierarchical cosmological framework}
%-------------------------------------------------------------

The question we investigate in this section is: assuming 
a modest  self-interacting cross section with 
$(\sigma/m_x) \cdot v_{100}=10^{-24}$ cm$^2$ GeV$^{-1}$ are  halo soft
cores produced by collisional dark matter in {\it a cosmological framework} 
stable or can they collapse? 
In other words, can dark halo density profiles
be soft at the centre and changing slowly their slope 
reaching the NFW shape or  steeper, due to the collision, in a Hubble time?

\begin{figure}
\epsfig{file=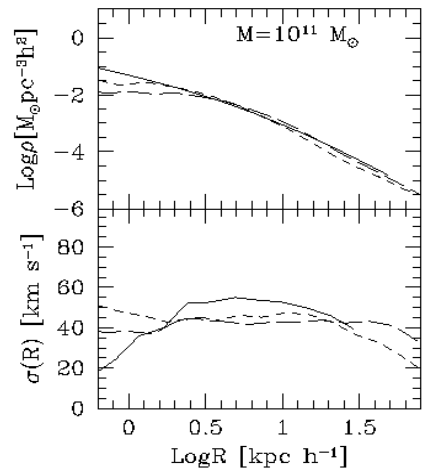,angle=0,width=10cm,height=12cm}
\@
\vskip-2cm
\caption{In the top panel  the dark matter 
density profiles of a halo of 
M$=10^{11} \ h^{-1} M_{\odot}$ are shown as obtained in a collisional CDM scenario.
The halo have been modelled assuming a energy dependent cross:
 $(\sigma/m_x)\cdot v_{100}=10^{-24}$ cm$^2$ GeV$^{-1}$.
Short dashed lines are the halo density profile
after 7.5 Gyr, while  long-dashed lines are the halo mass density 
distribution after a Hubble time. In the same plot the solid
line is the NFW profile.
In the bottom panel the corresponding 
halo radial dispersion velocity
profiles are shown.}
\end{figure}

Using the N-body technique described above we run 
simulations for a galactic halo and a cluster halo.
In Figure 5 we show the dark matter 
density profiles of a halo of 
M$=10^{11} \ h^{-1} M_{\odot}$ as obtained in a collisional CDM model.
The halo have been modelled assuming a energy dependent cross 
section with the same value as above. 
In the top panel the short dashed line shows the halo density profile
after 7.5 Gyr, while the long-dashed line is the halo mass density 
distribution after a Hubble time. In the same plot the solid
line is the NFW profile.\\ 
{\it Note that within a Hubble time 
a modest self-interaction cross section value is able to create
soft cores. No core collapse occurs within a Hubble time.} 

\begin{figure}
\epsfig{file=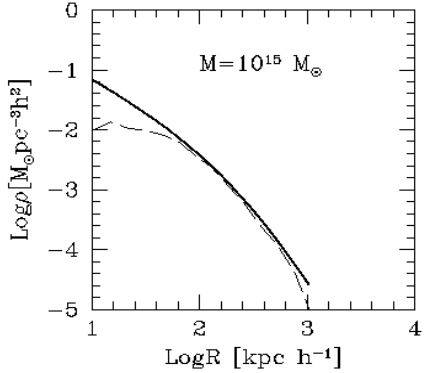,angle=0,width=10cm,height=12cm}
\@
\vskip-2cm
\caption{The dark density profile is shown
for a halo of $10^{15} \ h^{-1} M_{\odot}$, using a energy independent 
cross section: 
$(\sigma/m_x)\cdot v_{100} \sim 10^{-24}$ cm$^2$ GeV$^{-1}$. 
The long-dashed line
is the halo density profile after one Hubble time and the solid line
is the NFW profile.}
\end{figure}

The bottom panel of Figure 4 shows the corresponding 
halo radial dispersion velocity
profiles. The
velocity distribution of the particles reaches roughly a constant velocity 
dispersion under the effects of collisions resulting in a central 
non-singular isothermal density profile. 

In Figure 6 the dark density profile is shown
for a halo of $10^{15} \ h^{-1} M_{\odot}$. The long-dashed line
is the halo density profile after one Hubble time and the solid line
is the NFW profile.
Even in this case the halo does not undergo the core catastrophe
within a Hubble time.
We interpret this result assuming that  this is the case in which 
any trend towards the gravothermal catastrophe
is avoided by the competition between a mass aggregation 
determined by a {\it continuous}  halo merging history 
and a thermalization process
by collisions. In fact we assign to the particles cosmological
initial conditions assuming that the halo merging history
is a continuous process.

\begin{figure}
\epsfig{file=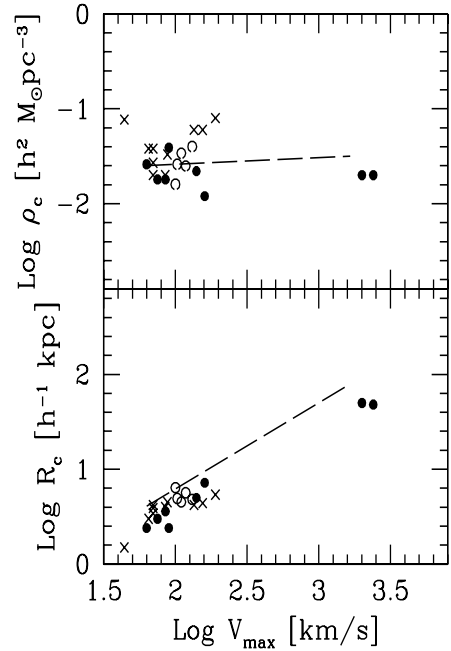,angle=0,width=10cm,height=12cm}
\@
\vskip-2cm
\caption{The halo central densities and the core radii predicted
by the models are compared with those inferred from HI rotation curves
of dwarf and LSB galaxies (diagonal crosses), H$\alpha$ 
rotation curves from Swaters et al. (2000), taking into
account the halo adiabatic contraction (open circles) and 
H$\alpha$ rotation curves  of LSB and dwarf galaxies (filled circles) 
by Marchesini et al. (2002).
The filled circles on the right of the panels are two galaxy 
clusters with evidence of soft cores: Cl 0024+1654 
and Abell 1795. Dashed lines are the halo central densities and core radii
as a function of the maximum circular velocity 
predicted by self-interacting dark matter with $(\sigma/m_x)\cdot v_{100}
\simeq 10^{-24}$ cm$^2$ GeV$^{-1}$.}
\end{figure}

In Figure 7, the central densities and the core radii predicted
by the models are compared with those inferred from HI rotation curves
of dwarf and LSB galaxies (diagonal crosses), H$\alpha$ 
rotation curves from Swaters et al. (2000), taking into
account the halo adiabatic contraction (open circles) and 
H$\alpha$ rotation curves  of LSB and dwarf galaxies (filled circles) 
by Marchesini et al. (2002).
The filled circles on the right of the panels are two galaxy 
clusters with evidence of soft cores: Cl 0024+1654 
and Abell 1795.
The dashed lines are the predictions of the model
for a cross section $(\sigma/m_x)\cdot v_{100}=10^{-24}$ cm$^2$ GeV$^{-1}$.

A feature of self-interacting dark matter, with any cross section
is to produce halos that are more spherical than the CDM halos, due to the 
isotropic nature of collisions.
Thus, collisions tend to isotropize the velocity ellipsoid. Consequently
the shapes of self-interacting dark matter halos and CDM halos should 
be different and provide an interesting observational constraint to
discriminate the two scenarios. Miralda-Escud\'{e} (2002), providing
lensing data for the galaxy cluster MS2137-23, shown that at a  radius $R\approx 70$ kpc,
anisotropies start in the velocity ellipsoid, constraining the cross section 
between dark particles to be less than 0.02 cm$^2$g$^{-1}$ in order to create
anisotropies in agreement with the observations.

Dav\'{e} et al. (2001) studied the the axisymmetry and the flattening for galactic
halos using N-body simulations with constant cross section between dark particles.
After the completition of this work, a study of N-body simulations of 
self-interacting dark halos appeared by Col{\'\i}n et al. 2002, in which the 
authors analyse the ellipticities in the halos, even using a cross section inversely
proportional to the halo dispersion velocity.
Due to the limitations of our method, cosmological triaxiality and a comparison
to the lensing data  cannot be
developed within our approach. Hovewer the effects of the self-interaction on an elongated
structure artificially introduced in the early evolution of the halo have
been explored. 
We define the axisymmetry $q$ and  the flattening $s$, following the same prescription used 
in Dav\'{e} et al. (2001). A inertia tensor is defined:\\
\begin{equation}
M_{ij}=\Sigma \frac{x_i x_j}{r^2}; \ \ \ \ r^2\equiv \Big(x_1^2+\frac{x_2^2}{q^2}+\frac{x_3^2}{s^2}\Big) 
\end{equation}
with the sum is over all particles with coordinates ($x_1,x_2,x_3$) and distance $r$ from the halo centre
and $s$ and $q$ the axis ratio:\\
\begin{equation}
q=\Big(\frac{M_{yy}}{M_{xx}}\Big)^{1/2}; \ \ \ s=\Big(\frac{M_{zz}}{M_{xx}}\Big)^{1/2}
\end{equation}
with $M_{xx},M_{yy},M_{zz}$ are the eingenvalues of the inertia tensor $M$.  

Since the observational constraints on anisotropies
existence come from galaxy cluster we feel more interesting to analyse the galaxy cluster scale 
halo with M=$10^{15} \ h^{-1} M_{\odot}$ and to assume the same cross section
capable to reproduce the soft core in agreement with the observations of Cl0024+1654,
$(\sigma/m_x)\cdot v_{100}\simeq 10^{-24}$ cm$^2$ GeV$^{-1}$.
Figure 8 shows
how the galaxy cluster halo is more spherical if a self-interacting scenario is working.
For a cluster with  $r_{vir}=1.5$ Mpc any early triaxiality desappears inside
a region of roughly  120 kpc (for h=0.65). This is not surprising because
particle collisions induced by self-interacting dark matter tend to recover the spherical symmetry. However
in realistic cases major mergers may be a continuous source for anisotropies, even 
within a few hundred kpc from the centre.

\begin{figure}
\epsfig{file=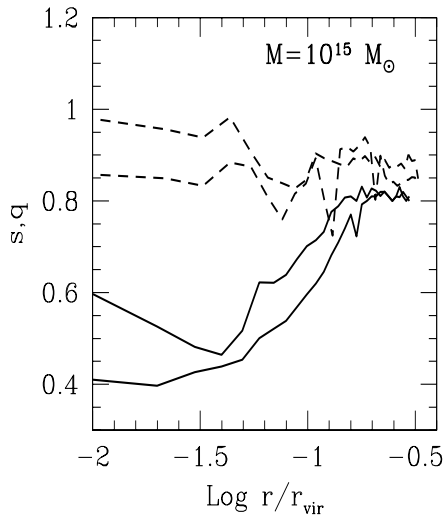,angle=0,width=10cm,height=12cm}
\@
\vskip-2cm
\caption{Axisymmetry $q$ and flattening $s$ for a cluster halo of M=$10^{15} \ h^{-1} M_{\odot}$
versus $r/r_{vir}$.
Solid lines represent $s$ (lower line) and $q$ (higher line) runs with collisionless 
dark matter halos. Dashed lines are simulations with collisional dark matter with
cross section $(\sigma/m_x)\cdot v_{100}\simeq 10^{-24}$ cm$^2$ GeV$^{-1}$.}
\end{figure}

%----------------------------------------------------------------
\section{Discussion and Conclusion}
%----------------------------------------------------------------

We have explored the dark halo density profiles in a  collisional dark
matter scenario studying isolated or cosmological halos of any size and mass.
This study presents two key points: {\it i)} we use 
a modified N-body code where the cosmological initial conditions
are assigned with a new approach; and  {\it ii)} we use a 
self-interacting cross section inversely proportional to the particle
collision velocity.

Our results may be summarized as follows:
\begin{itemize}
\item  Concerning the dynamical study of isolated halos we 
 have followed the evolution of different initial profiles:
 the Hernquist model and the King profile. 
 Surprisingly, the virialized  halo density profiles for all
 models undergo  core collapse induced by collisional dark matter.
 Once the core size is reduced, core collapse is a very rapid
process. 
 The shape of the cross section (constant or dependent upon the particle collision velocity)
 does not affect core collapse. Our N-body simulations
performed with a cross section dependent on the particle
velocity dispersion  produce similar
results to the N-body simulations by 
Burkert (2000) and Kochanek $\&$ White (2000) obtained with a constant 
cross section. 

\item In the hierarchical framework, the halo core catastrophe 
is avoided from the balance between mass aggregation process
induced by gravitation and thermalization process induced by
collisions. If the balance fails, collisions between particles
induce core collapse in a short time. Large cross section
values cause even faster core collapse.

\item If the evidence of soft cores is accepted for Cl 0024+1654
 and Abell 1795, the observed central densities of both clusters
and dwarf galaxies  are reproduced assuming a cross section value of:
$(\sigma/m_x)\cdot v_{100} \approx 10^{-24}$ cm$^2$ GeV$^{-1}$.
The shape and the value of the cross section suggested by our work 
are in the range of values in which  the core collapse 
and the evaporation problem are avoided (Gnedin $\&$ Ostriker 2001; Hennawi $\&$ Ostriker 2002).
\end{itemize}
Observations of the mass distribution at the centre of galaxy clusters
are crucial for the debate regarding  the nature of the dark matter. If future
observations cannot confirm the evidence for soft cores on galaxy cluster
scales, then the soft core question remains confined to dwarf  
and LSB galaxies. In this case 
different  mechanisms could play a role in producing soft cores in
dark matter dominated galaxies.

\section*{Acknowledgments}
We wish to thank the anonymous referee for his comments and suggestions that 
have improved this work. 
We are grateful to Hugh Couchman for having made available his adaptive 
P$^3$M-SPH code HYDRA. ED thanks Henry Lee for an early reading of 
the paper.

\bsp

\label{lastpage}

\end{document}